# CH-π interaction-induced deep orbital deformation in a benzene-methane weak binding system

Jianfu Li, Rui-Qin Zhang[*]

*Department of Physics and Materials Science, City University of Hong Kong, Hong Kong SAR, China*

**Abstract**

The nonbonding interaction between benzene and methane, called CH-π interaction, plays an important role in physical, chemical, and biological fields. CH-π interaction can decrease the system total energy and promote the formation of special geometric configurations. This work investigates systemically the orbital distribution and composition of the benzene-methane complex for the first time using *ab initio* calculation based on different methods and basis sets. Surprisingly, we find strong deformation in HOMO-4 and LUMO+2 induced by CH-π interaction, extending the general view that nonbonding interaction does not cause orbital change of molecules.

**Introduction**

In the last two decades, the benzene-methane complex has been used as a model system to study CH-π interaction, which is considered a weak interaction and has been found to play important roles in the physical, chemical, and biological properties of a variety of substances [1–5]. An electronic spectroscopy study demonstrated that a benzene-methane cluster has an on-top structure where the methane molecule is located on the $C_6$ axis of the benzene moiety to hold the $C_3$ or a higher symmetry [6]. High-level *ab initio* calculations of the benzene-methane cluster confirmed that the on-top type isomer indicated by the electronic spectroscopy is the most stable structure, with 1.45 kcal/mol binding energy at the CCSD(T)-level calculation [7]. The zero-point energy corrected binding energy (1.13 kcal/mol) agrees well with the experimental value (1.03-1.13 kcal/mol) [8].

---
[*] Corresponding author.
   *E-mail address:* aprqz@cityu.edu.hk (Rui-Qin Zhang).

In the benzene-methane complex, the dispersion energy contributes most in typical cases, where the CH bond is the hydrogen donor and the $C_6H_6$ ring is the CH-acceptor [9,10]. CH-π interaction not only reduces the total energy and forms a special configuration of the complex but also changes the properties of the CH and/or π group. In 1955, Pinchas first reported the blue shift of the C-H stretching frequency in the IR spectra of O-substituted benzaldehydes [11]. This phenomenon has since been reported in many experimental [12–14] and theoretical studies [15–18]. Besides the C-H stretching shift effect, CH-π interaction also can change the electronic structures of a π group. Li and Chen revealed that considerable band gaps can be opened in the band structures of graphene/$CH_4$ and silicene/silicane bilayers due to the broken equivalency of two sublattices inducted by XH/π (X = C, Si) interactions [19]. A common view is that nonbonding interaction does not involve orbital interaction due to the large distance (> 3 Å) between molecules. To date, there has been no research addressing the impact of CH-π interaction on molecular orbitals (MOs).

In the present article, for the first time, we systematically examine the MOs of the benzene-methane complex using *ab initio* calculations. The results reveal that the deformation of MOs occurs not only in deep occupied HOMO-4 but also in unoccupied LOMO+2.

**Computational details**

The *ab initio* calculations of structural relaxations and electronic properties were carried out with the Gaussian 09 package [20]. A long-range corrected hybrid density functional ωB97X [21] was used to explore the configuration and interaction energy in the ground state. The hybrid density functional ωB97X includes 100% long-range exact exchange, applying generalized gradient expressions for short-range exchange. The basis set superposition error (BSSE) [22] was corrected for all calculations using the counterpoise method [23,24]. Different basis sets, including 6-311G, 6-311G*, 6-311G**, cc-pVDZ, and cc-pVTZ, were used to explore the basis set effect. Single point calculations were used to show the impact on orbitals of intermolecular interaction based on ωB97X methods.

**Results and discussion**

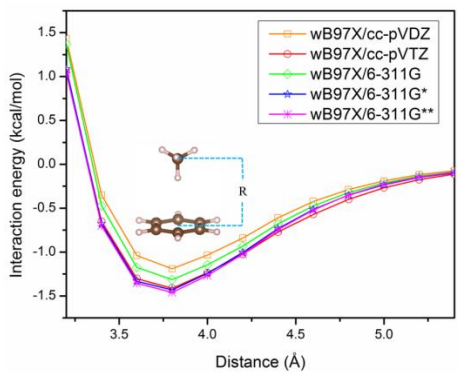

Figure 1. The BSSE-corrected intermolecular interaction energy of the benzene-methane system considered, calculated using the ωB97X method.

As many experimental and theoretical results have proven that the benzene-methane complex has an on-top type isomer configuration [6–8], we took this configuration as our model, as shown in the middle of Figure 1. The BSSE-corrected intermolecular interaction energies of the benzene-methane complex, calculated at the ωB97X level using 6-311G, 6-311G*, 6-311G**, and cc-pVXZ (X = D and T) to evaluate the basis set effect, are provided in Figure 1. The BSSE-corrected interaction energy is -1.46 kcal/mol with an intermolecular distance R equal to 3.8 Å using 6-311G** basis sets, which is in good agreement with previous results at the CCSD(T) level with basis set limit [7]. Using this configuration as the input structure, the BSSE-corrected intermolecular interaction energy is -1.51 kcal/mol after relaxation, and the corresponding distance R is 3.754 Å.

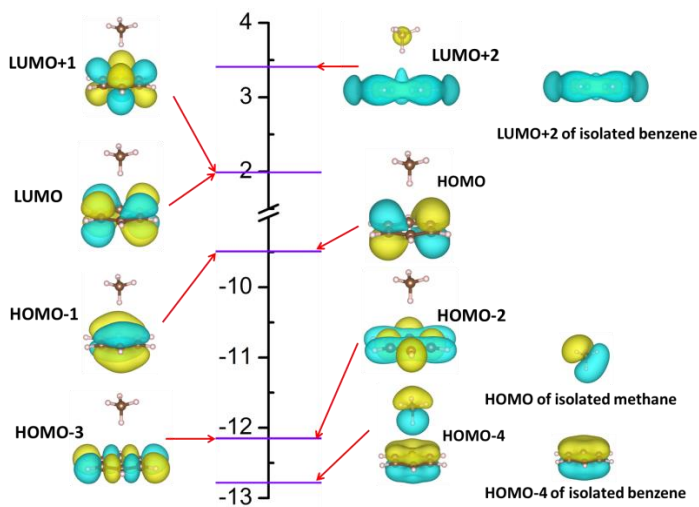

Figure 2. The front molecular orbitals of the benzene-methane complex calculated by ωB97X/6-311G** basis sets. The isosurface value is ±0.03 and the blue and yellow coloration denote negative and positive value, respectively. Relative orbitals of isolated benzene and methane have also been plotted.

The orbital distributions of the benzene-methane complex are plotted in Figure 2 together with the corresponding energy levels. The results indicate that the orbital distribution of the benzene-methane complex is not a simple superposition of benzene and methane. As shown in Figure 2, from HOMO-3 to LUMO+1, the contributions only come from benzene. However, for HOMO-4 and LUMO+2, the orbitals are localized not only in benzene but also in methane. In other words, HOMO-4 and LUMO+2 are formed by both benzene and methane under CH-π interaction. Compared with the LUMO+2 of the isolated benzene, the part of LUMO+2 localized in the benzene of the complex has an obvious peak pointing to methane and loses its symmetry about the benzene plane. HOMO-4 consists of a delocalized π orbital of benzene and a σ orbital of methane. The distribution of the π orbital of benzene becomes smaller and the center has an obvious depression compared with the HOMO-4 in isolated benzene. The σ orbital of methane changes its orientation to the center of benzene relative to the HOMO of isolated methane. Moreover, the plots of the corresponding MOs of the interacting monomers at a longer distance (10 Å) as shown in Figure A1 clearly show that the mixture of MOs disappears, which further confirms that the above described orbital deformation is induced by the CH-π interaction.

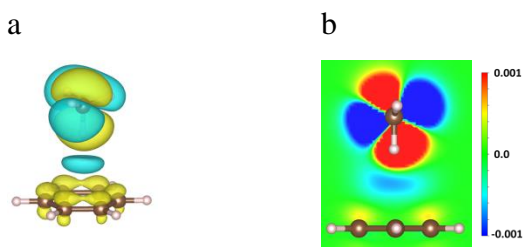

Figure 3. a) and b) are the isosurface and the cross section of partial charge density difference corresponding to HOMO-4 of the benzene-methane complex, respectively. The blue region represents the electron donor and the yellow region represents the electron acceptor. The isosurface value is ±0.0005.

In occupied HOMO-4, the deformation of the orbital will cause a change in partial charge density, as shown in Figure 3. The calculated results indicate that the partial charge distribution of both benzene and methane has been changed under CH-π interaction. For benzene within the complex, the deformation of HOMO-4 results in a corresponding charge accumulating around carbon atoms. As for methane, the charge transfers in the direction of the symmetry axis of the complex.

To give a detailed description of the contribution of benzene and methane and confirm the above results, the orbital composition of the benzene-methane complex is listed in Table 1 based on the ωB97X/6-311G** calculations. The results show clearly that HOMO-4 and LUMO+2 are contributed to by both benzene and methane. From HOMO-3 to LUMO+1, the contribution of methane is less than 1%. However, methane contributes 52.46% to HOMO-4 and 19.23% to LUMO+2. Under CH-π interaction, the frontier MOs of the complex, from HOMO-3 to LUMO+1, are composed completely by benzene, the deep (HOMO-4) and out (LUMO+2) MOs do not consist of benzene or methane only, but of both.

Table 1. Orbital composition of the benzene-methane complex.

| | HOMO-4 | HOMO-3 | HOMO-2 | HOMO-1 | HOMO | LUMO | LUMO+1 | LUMO+2 |
|---|---|---|---|---|---|---|---|---|
| $CH_4$ | 52.46% | 0.00% | 0.00% | 0.20% | 0.20% | 0.54% | 0.55% | 19.23% |
| $C_6H_6$ | 47.54% | 100.00% | 100.00% | 99.80% | 99.80% | 99.46% | 99.45% | 80.77% |

**Conclusions**

We have revealed for the first time the deformation of MOs of the benzene-methane complex, a representative weak interaction system. Our *ab initio* calculations based on a well-demonstrated density functional method, ωB97X, and various basis sets, provide the details of the distribution and composition of MOs and the partial charge density difference of HOMO-4. From HOMO-3 to LUMO+1, the MOs of the benzene-methane complex are composed of atomic orbitals of benzene. However, HOMO-4 and LUMO+2 are composed of considerable contributions from both benzene and methane. The deformation of HOMO-4 results in a charge rearrangement relative to the corresponding MOs in isolated molecules. Our results demonstrate that CH-π interaction between benzene and methane not only reduces the total energy and forms a special configuration of the complex, but also affects the distribution and composition of the MOs. This finding enriches our understanding of weak interaction systems.

**Acknowledgements**

The work described in this paper was supported by a grant from the Research Grants Council of Hong Kong SAR (Project No. CityU 103913) and by the High Performance Cluster Computing Centre, Hong Kong Baptist University, which receives funding from the Research Grant Council, the University Grant Committee of HKSAR, and Hong Kong Baptist University.

**Supporting information**

| Labels of orbitals | $R = 3.8$ Å | $R = 10$ Å | Labels of orbitals | $R = 3.8$ Å | $R = 10$ Å |
|---|---|---|---|---|---|
| LUMO+2 | | | HOMO-1 | | |
| LUMO+1 | | | HOMO-2 | | |
| LUMO | | | HOMO-3 | | |
| HOMO | | | HOMO-4 | | |

**Figure A1**. Front occupied orbitals of benzene-methane complex. R is the intermolecular distance between carbon atom of methane and center of benzene.